\documentclass[10pt,conference]{IEEEtran}
\IEEEoverridecommandlockouts
\usepackage{hyperref}
\usepackage{multirow}
\usepackage{cite}
\usepackage{amsmath,amssymb,amsfonts}
\usepackage[capitalise]{cleveref}
\usepackage{algorithmic}
\usepackage{textcomp}
\usepackage{xcolor}
\usepackage{graphicx}
\usepackage{booktabs}

\def\BibTeX{{\rm B\kern-.05em{\sc i\kern-.025em b}\kern-.08em
    T\kern-.1667em\lower.7ex\hbox{E}\kern-.125emX}}

\newcommand{\revision}[1]{\textcolor{black}{#1}}

\newcommand{\tool}{DesignRepair}

\usepackage[most]{tcolorbox}
\newtcolorbox{myquote}{
  colback=gray!20,   
  colframe=black!75!white,  
  fonttitle=\bfseries,       
  left=1em, right=1em,       
  boxrule=0pt,            
  arc=8pt,                   
  auto outer arc,            
  lifted shadow={1mm}{-1mm}{2mm}{0.3}  
}

\usepackage{subcaption}

\begin{document}

\title{\tool{}: Dual-Stream Design Guideline-Aware Frontend Repair with Large Language Models
}

\author{
    \IEEEauthorblockN{ 
    Mingyue Yuan\IEEEauthorrefmark{2}\IEEEauthorrefmark{3},
                       Jieshan Chen\IEEEauthorrefmark{3}\IEEEauthorrefmark{5}\IEEEauthorrefmark{1},
                       Zhenchang Xing\IEEEauthorrefmark{3}\IEEEauthorrefmark{6},
                       Aaron Quigley\IEEEauthorrefmark{2}\IEEEauthorrefmark{3},
                       Yuyu Luo\IEEEauthorrefmark{4},
                       Tianqi Luo\IEEEauthorrefmark{4}}
    \IEEEauthorblockN{ 
                       Gelareh Mohammadi\IEEEauthorrefmark{2},
                       Qinghua Lu\IEEEauthorrefmark{2}\IEEEauthorrefmark{3},
                       Liming Zhu\IEEEauthorrefmark{2}\IEEEauthorrefmark{3}
                      }
    \IEEEauthorblockA{
    \IEEEauthorrefmark{2}University of New South Wales, Australia}
     \IEEEauthorblockA{
    \IEEEauthorrefmark{3}CSIRO’s Data61, Australia}
     \IEEEauthorblockA{
    \IEEEauthorrefmark{4}The Hong Kong University of Science and Technology (Guangzhou), China} 
     \IEEEauthorblockA{
    \IEEEauthorrefmark{5} Institute for Advanced Study,
Technical University of Munich, Germany}
     \IEEEauthorblockA{
    \IEEEauthorrefmark{6}Australian National University, Australia\\
    Email: \IEEEauthorrefmark{2}\{mingyue.yuan, g.mohammadi\}@unsw.edu.au, \IEEEauthorrefmark{4}\{yuyuluo, tluo553\}@hkust-gz.edu.cn
    \\ \IEEEauthorrefmark{3}\{jieshan.chen, zhenchang.xing, a.quigley, qinghua.lu, liming.zhu\}@data61.csiro.au 
    }
    \thanks{\IEEEauthorrefmark{1} Jieshan Chen is the corresponding author.}
    \thanks{This work is partially funded by the Dieter Schwarz Foundation and the Technical University of Munich – Institute for Advanced Study, Germany.}
}

\maketitle

\begin{abstract}
The rise of Large Language Models (LLMs) has streamlined frontend interface creation through tools like Vercel's V0, yet surfaced challenges in design quality (e.g., accessibility, and usability). Current solutions, often limited by their focus, generalisability, or data dependency, fall short in addressing these complexities. Moreover, none of them examine the quality of LLM-generated UI design.
In this work, we introduce \tool{}, a novel dual-stream design guideline-aware system to examine and repair the UI design quality issues from both code aspect and rendered page aspect. We utilised the mature and popular Material Design as our knowledge base to guide this process. Specifically, we first constructed a comprehensive knowledge base encoding Google's Material Design principles into low-level component knowledge base and high-level system design knowledge base. After that, \tool{} employs a LLM for the extraction of key components and utilizes the Playwright tool for precise page analysis, aligning these with the established knowledge bases. Finally, we integrate Retrieval-Augmented Generation with state-of-the-art LLMs like GPT-4 to holistically refine and repair frontend code through a strategic divide and conquer approach.
Our extensive evaluations validated the efficacy and utility of our approach, demonstrating significant enhancements in adherence to design guidelines, accessibility, and user experience metrics. 
\end{abstract}


\section{Introduction}

The emergence of advanced large language models (LLMs), such as GPT-4~\cite{bubeck2023sparksgpt4} and Meta's Llama~\cite{r2024codellama}, has catalyzed transformative changes across various fields, including legal~\cite{guha2023legalbench, koreeda2021contractnli, hendrycks2021cuad, wang2023maud}, finance~\cite{cheng2024adapting}, and software development~\cite{hou2024large}, unlocking tremendous new possibilities.  
For frontend software development, commercial and open-source tools, such as Vercel's V0\cite{v0dev2024}, Imagica\cite{Imagica2024} and OpenV0~\cite{raidendotai_openv0_2024}, have attracted considerable interest using the capabilities of LLMs. 
These applications adeptly transform user-provided textual or visual prompts into concrete, well-structured high-fidelity user interfaces, along with their associated front-end code, providing substantial support to designers and developers. To augment the quality of the generated UI and its accompanying code, these tools commonly integrate with established UI design libraries (e.g., HTML, CSS, JS frameworks, and libraries), including popular ones like React~\cite{react_dev_2024}, Shadcn~\cite{shadcn_ui_2024} and Tailwind CSS~\cite{tailwindcss_2024}. These libraries serve as foundational knowledge bases, empowering these tools to generate aesthetically pleasing and technically sound designs. 

\begin{figure}[t]
    \centering
    \begin{subfigure}[b]{0.23\textwidth}
        \centering
        \includegraphics[width=\linewidth]{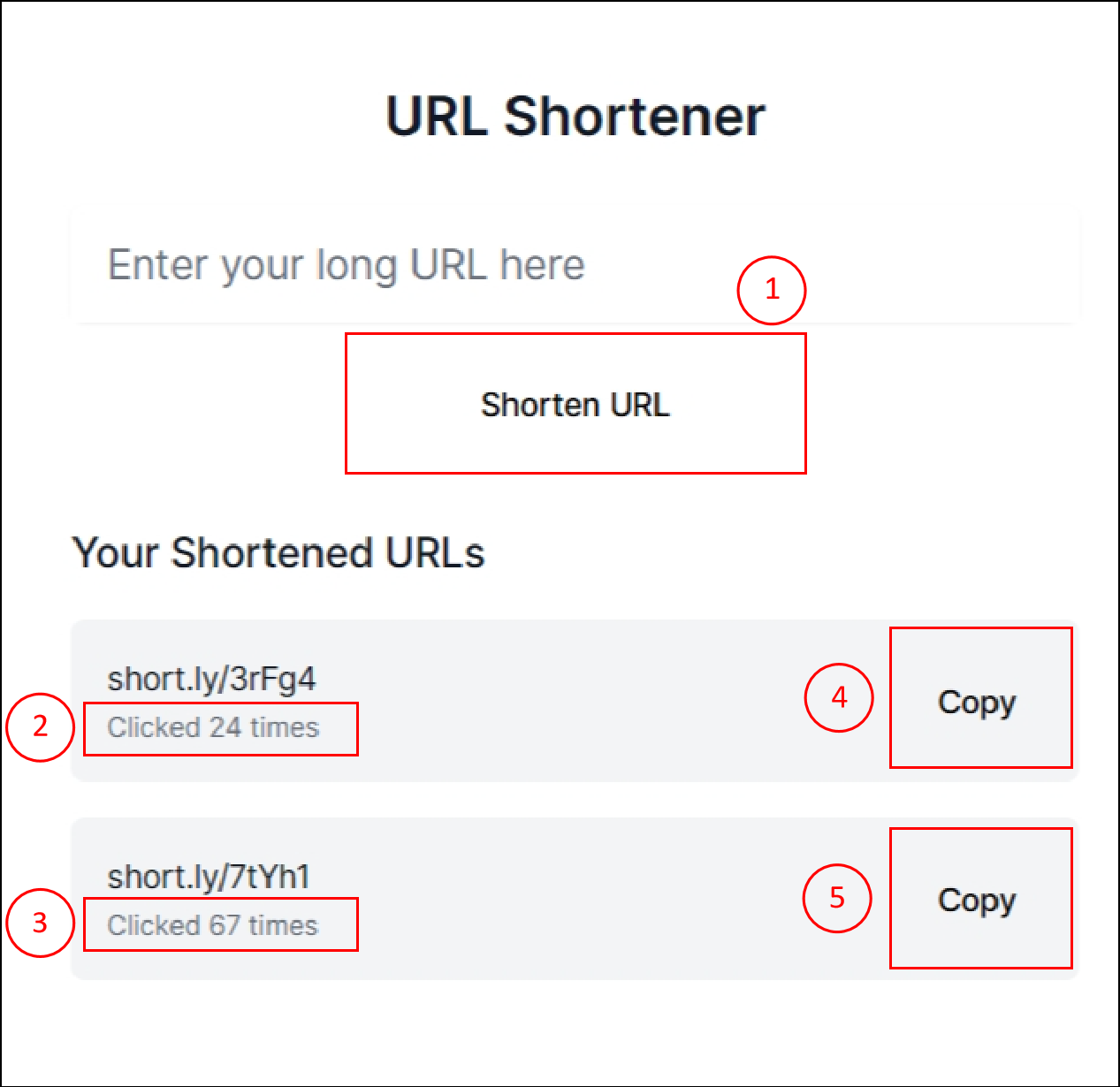}
        \captionsetup{font=scriptsize}
        \caption{Initial UI Displaying Text and Button Color Contrast Issues, Highlighted in Red}
        \label{fig:exp1}
    \end{subfigure}
    \hfill 
    \begin{subfigure}[b]{0.23\textwidth}
        \centering
        \includegraphics[width=\linewidth]{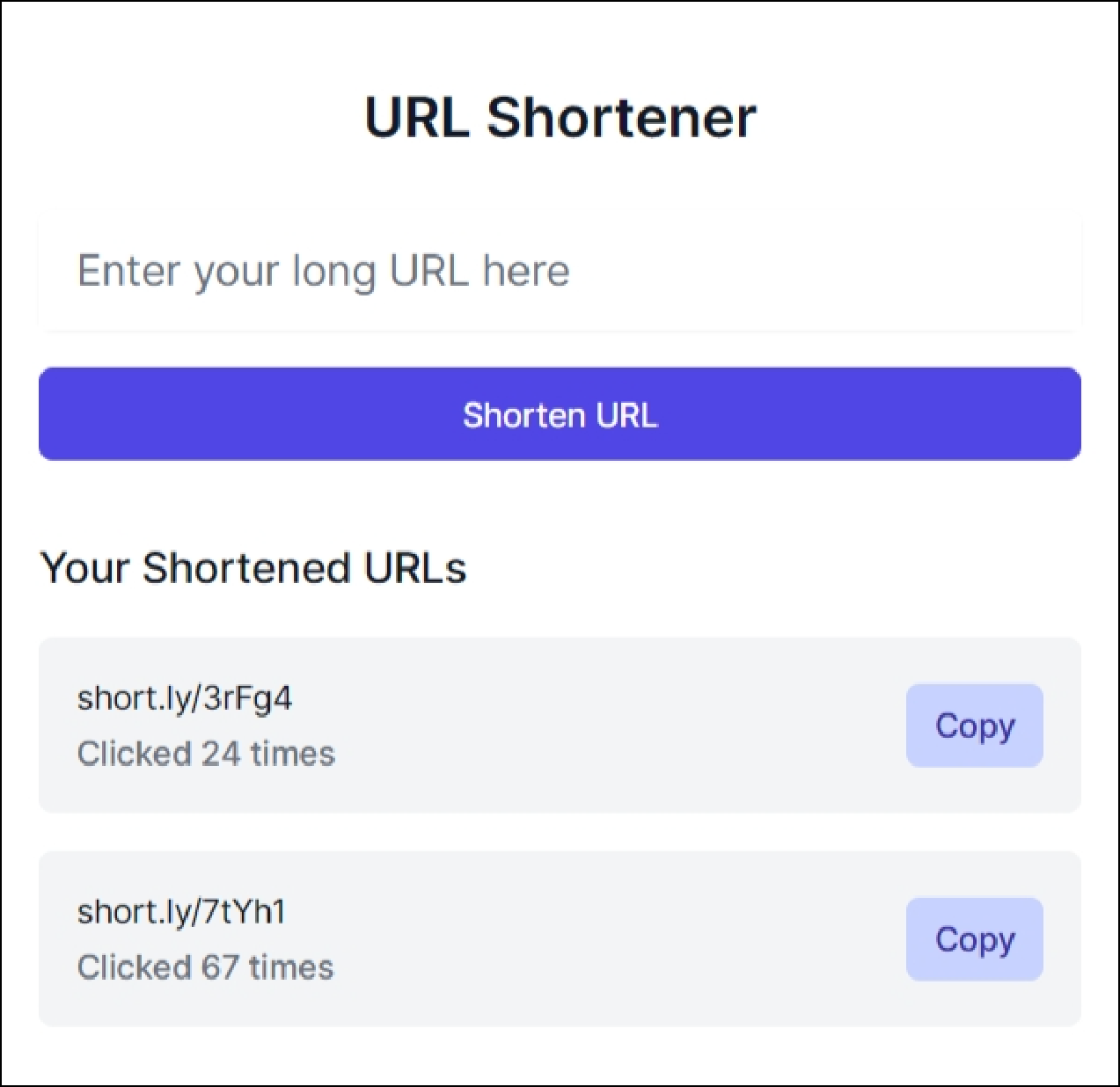}
        \captionsetup{font=scriptsize}
        \caption{Repaired UI using our method, showcasing resolved issues}
        \label{fig:exp2}
    \end{subfigure}
    \caption{``URL Shortener'' by Vercel's V0: Initial Design vs. Improved Design. 
    }
    \vspace{-8mm}
\end{figure}

Despite their advanced capabilities, these tools continue to face two significant challenges.
Firstly, while they are adept at converting prompts into well-structured UI elements, their compliance with established design principles, such as material design~\cite{material_design_3_2024}, often falls short (see \cref{sec:experiments}). This inadequacy is critical, as strict adherence to these principles is fundamental to crafting effective, accessible, and user-friendly interfaces. Such high-quality interfaces play a vital role in setting an application apart from its competitors, boosting user downloads, reducing user complaints, and enhancing user retention~\cite{liu2020owl, chen2022towards, zhao2020seenomaly, yang2021don, chen2020wireframe, vajjala2024motorease}. 
Secondly, these tools exhibit limitations in the repair and refinement of generated designs. While they allow for additional prompts for UI adjustments, they often rely on users to manually identify and describe needed changes. This process can be cumbersome, requiring extensive labeling and in-depth knowledge~\cite{zhao2020seenomaly, yang2021don, Moran_2018}.
\cref{fig:exp1} shows an example. When prompted with ``URL shorten Chrome extension,'' Vercel's V0 successfully generated an appealing user interface. Nonetheless, upon closer examination, it exhibited five distinct design issues related to contrast and color in text and button elements, marked in red.

Many studies have been proposed to assess different facets of UI design development to ensure its quality, encompassing accessibility~\cite{microsoft_playwright_2024, googlechrome_lighthouse_2024, dequelabs_axe_core_2024, Liu_2020,mahmud2024using, salehnamadi2021latte}, tapability~\cite{swearngin2019modeling}, animation effects~\cite{zhao2020seenomaly}, and broader design principles and concerns~\cite{yang2021don, chen2023unveiling, liu2020owl}. 
Industrial tools like Playwright~\cite{microsoft_playwright_2024} and Google Lighthouse~\cite{googlechrome_lighthouse_2024} primarily address accessibility concerns, such as text contrast for readability. 
Swearing et al.\cite{swearngin2019modeling} utilized a classifier-based technique to identify discrepancies between the perceived and intended tappability of UI elements.
Yang et al.~\cite{yang2021don} introduced UIs-Hunter, a rule-based method for detecting deviations from material design principles in UIs.
Despite these advancements, they either tend to focus on specific, isolated issues within UI design, or they encounter challenges in generalizing to a broader spectrum of design aspects. 
\revision{Moreover, rule-based methods require extensive manual efforts which limits their usability and generalisability.} 


To tackle the aforementioned challenges and fill the existing research gaps, we introduce \tool{}, a novel knowledge-driven LLM-based approach specifically designed for enhancing the quality assurance and remediation of frontend UI and code. 
\tool{} adopts Material Design 3~\cite{material_design_3_2024} as the standard for evaluating quality and as the foundation of our knowledge base. 
To guarantee efficient and effective knowledge retrieval, catering to varying levels of repair granularity and diverse design requirements, we meticulously analyzed and restructured Material Design 3 into two organized knowledge bases: the component knowledge base and the system design knowledge base. The component knowledge base focuses on the detailed, fine-grained design principles of individual components, whereas the system design knowledge base provides broader guidance on the overall design perspective, ensuring coherence in elements such as layout and color harmony.

Given the frontend code and its corresponding rendered page, \tool{} initially extracts essential information, such as components and property groups, followed by retrieving relevant knowledge from our knowledge bases. This approach, considering both the source code and the user-perceived experience, provides a holistic and detailed examination.
Finally, \tool{} adopts a divide and conquer approach to iteratively repair the code. The process progresses from focusing on individual elements to considering the overall design, ensuring a comprehensive and systematic approach to repair frontend code.

We conducted an evaluation to measure the effectiveness and usefulness of \tool{}. \revision{We collected 196 design issues from AI-generated frontend code and 115 from GitHub projects consisting of UIs and their frontend code. Our evaluation shows that \tool{} effectively identifies and repairs design issues, achieving a recall of 89.3\% and precision of 86.6\% for AI-generated issues, and a recall of 85.2\% with precision of 90.7\% for GitHub issues. Additionally, our experiments on each step } highlights the effectiveness of the strategies we proposed.
Finally, a user study with 26 participants further confirms the high quality of our repairs. \revision{All our code and prompts are released at our GitHub repository~\footnote{\url{https://github.com/UGAIForge/DesignRepair}}.}

Our contributions are as follows:

\begin{itemize}
    \item We \revision{systematically analyzed} Material Design guidelines and converted them into structured knowledge bases, i.e., low-level component knowledge base and high-level system design knowledge base, enabling efficient and effective utilization of knowledge in design quality assurance processes.

    \item We introduced \tool{}, a novel dual-stream knowledge-driven, LLM-based method, uniquely designed to detect and repair design quality issues in frontend code. This approach simultaneously considers both the source code and the user-perceived rendered page, ensuring outstanding performance in design quality enhancement.

    \item We conducted extensive experiments and a user study with 26 participants, which demonstrate the effectiveness and usefulness of our approach.

\end{itemize}

\section{Background}

\subsection{Material Design Guidelines}
\label{sec:materialdesign}

\revision{Material Design 3~\cite{material_design_3_2024} is a broad design system built and supported by Google designers and developers, and can be implemented across Android, iOS, and web platforms.} 
Its latest version, Material Design 3, provides guidelines and examples to support important practices of user interface, ensuring personal, adaptive, and expressive experiences. 
It covers topics ranging from fine-grained design guidelines such as the application of individual components and color usage for specific meanings, to the high-level composition of components and common design patterns.

The guidelines consist of \textbf{three main sections}: foundations, styles, and components. 
\texttt{Foundations} inform ``the basis of any great user interface, from accessibility standards to essential patterns for layout and interaction.'' For example, it offers advice on simplifying the hierarchy of UI elements, making them more straightforward to perceive and understand.
\texttt{Styles} are ``the visual aspects of a UI that give it a distinct look and feel.'' For instance, they may include recommendations on color palettes to maintain a consistent appearance across the app's user interfaces, and guidelines for choosing colors for primary, secondary, and tertiary accent groups to enhance the user experience.
\texttt{Components} are ``interactive building blocks for creating a user interface. They can be organized into categories based on their purpose: Action, containment, communication, navigation, selection, and text input.'' These guidelines present best practices for selecting and utilizing components to effectively communicate the intended design purpose.
These three sections are interconnected, collectively providing a comprehensive framework for designing intuitive and visually cohesive user interfaces.

\begin{figure}[t]
    \centering
    \begin{subfigure}[b]{0.23\textwidth}
        \centering
        \includegraphics[width=\linewidth]{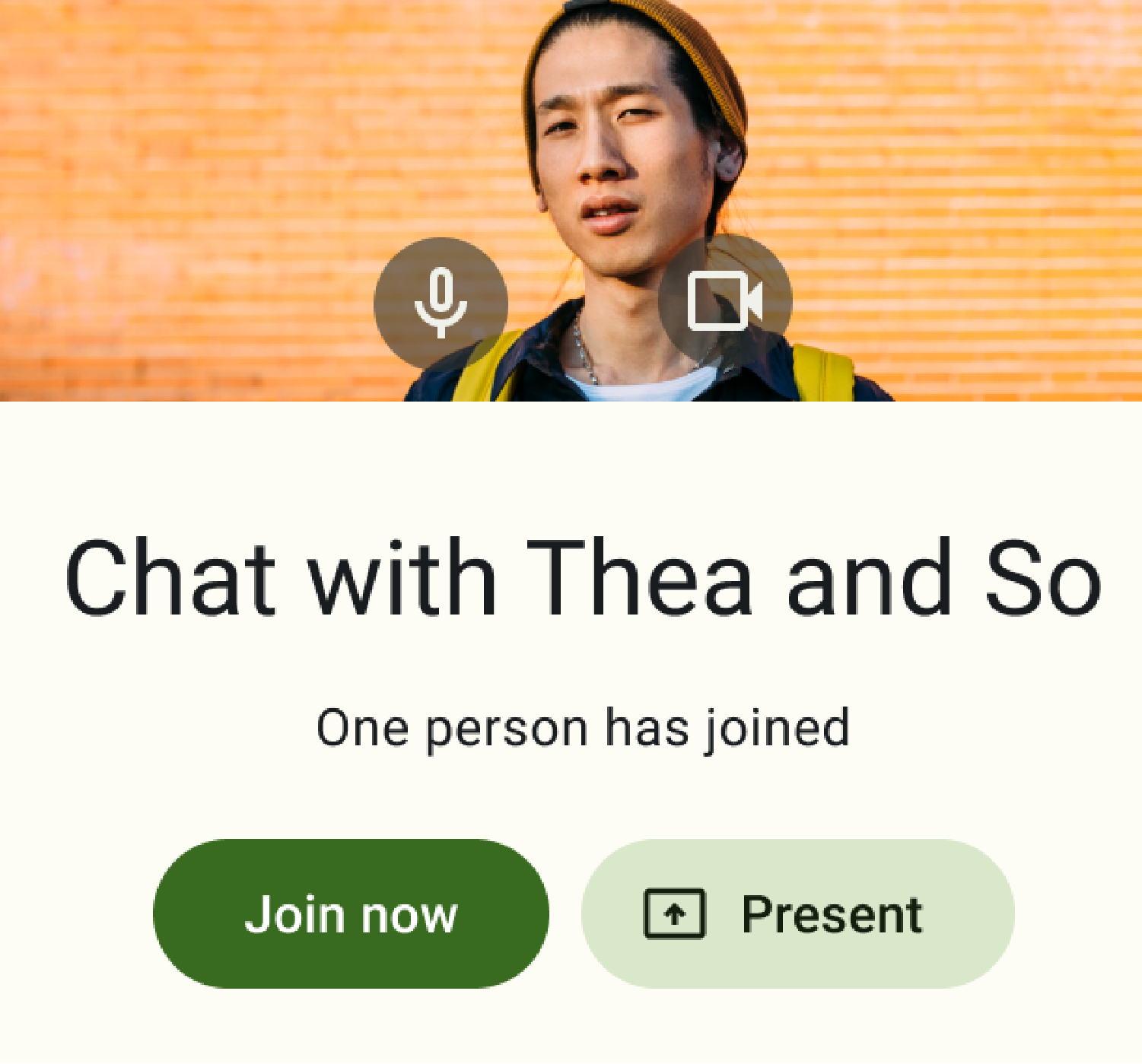}
        \captionsetup{font=scriptsize}
        \caption{\revision{\textit{\textbf{Recommended}}}: Use visually prominent filled buttons for the most important actions}
        \label{fig:exp2-soft}
    \end{subfigure}
    \hfill 
    \begin{subfigure}[b]{0.23\textwidth}
        \centering
        \includegraphics[width=\linewidth]{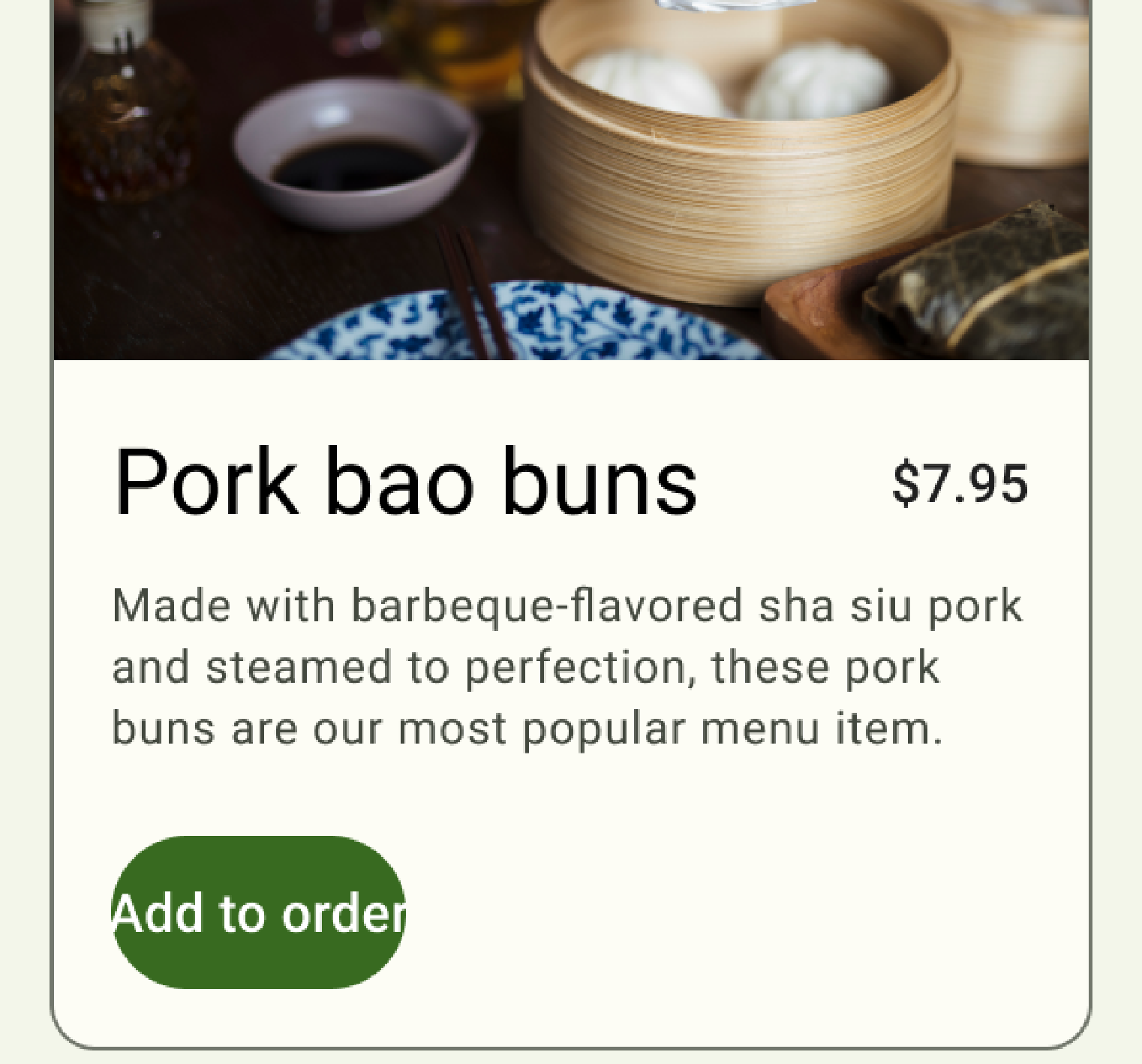}
        \captionsetup{font=scriptsize}
        \caption{\revision{\textit{\textbf{Do/Don't}}}: A button container's width shouldn't be narrower than its label text \\}
        \label{fig:exp2-hard}
    \end{subfigure}
    \caption{Examples of \textit{recommended} and \textit{don't} design guidelines in Material Design 3. Both guidelines are related to Button component in the \texttt{Components} section, and regard to the \textit{Usage} aspect. We categorised the \textit{recommended} guideline as a soft constraint, and \textit{do/don't} guidelines as hard constraints.}
    \vspace{-6mm}

\end{figure}

Within these three sections, they further classify the guidelines according to \textbf{various aspects of UI/UX design principles}, such as anatomy, behavior, responsive layout, usage, placement, and other categories. For instance, the \textit{Usage} aspect focuses on how and when to utilize a specific component, color, or layout.
Moreover, they utilize clear text and illustrative examples to explain \textbf{what to do (\textit{do}), what to avoid (\textit{don't}), and what is \textit{recommended}}. 
Adhering to the \textit{do} and \textit{don't} guidelines is crucial for delivering an optimal user experience; violating them can lead to user confusion or dissatisfaction. On the other hand, following the \textit{recommended} guidelines, while not mandatory, can enhance user engagement and satisfaction by refining the overall usability and aesthetic appeal of the interface.
For instance, \cref{fig:exp2-soft} in the \texttt{Components} section includes a \textit{recommended} guideline for the Button component, stating ``use visually prominent filled buttons for the most important action,'' illustrated with a ``Join now'' button example. 
On the other hand, \cref{fig:exp2-hard} offers a \textit{don't} guideline, ``Don’t: A button container's width shouldn't be narrower than its label text. '' shown with an example where a button's label exceeds its container width, leading to a poor user experience. These two examples both focus on the \textit{Usage} aspect of the Button component.

\section{Approach}

\begin{figure}
    \centering
    \includegraphics[width=0.9\linewidth]{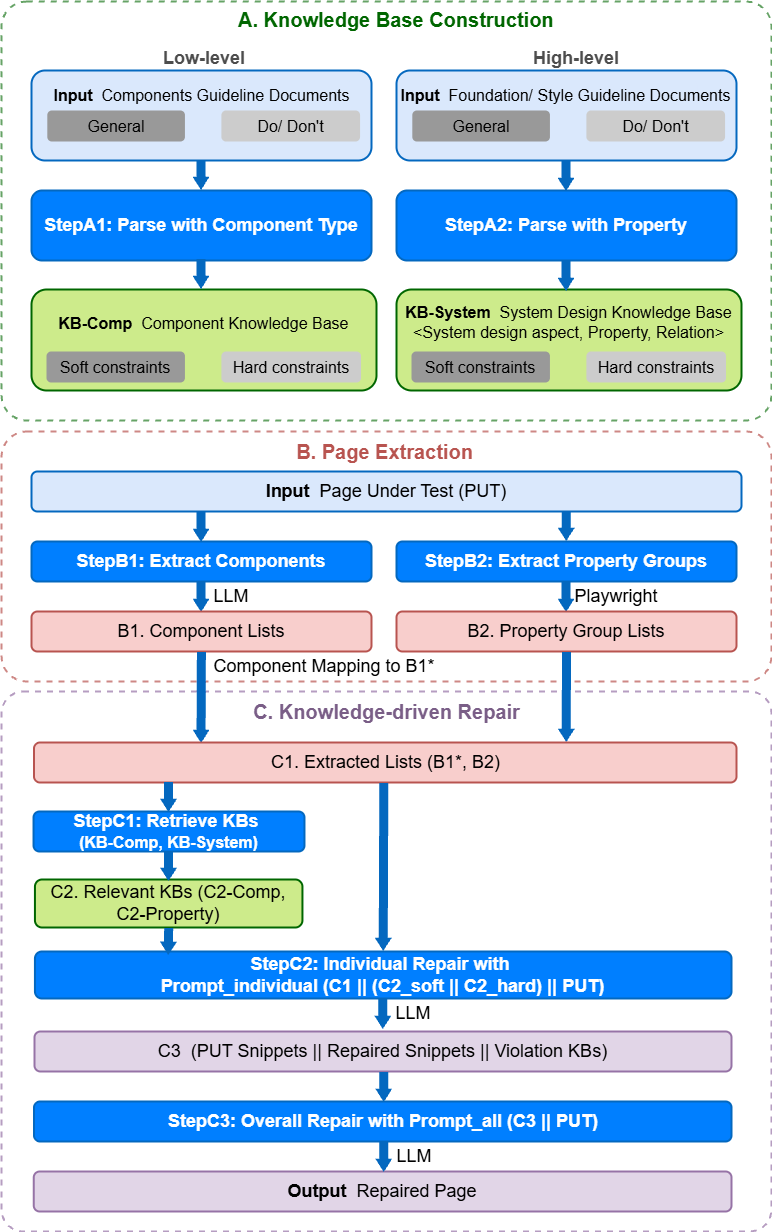}
    \caption{\revision{Framework Overview of Our Approach}
    }
    \label{fig:overview}
    \vspace{-6mm}
\end{figure}

\cref{fig:overview} illustrates the overview of our approach, \tool{}, which consists of three phases, namely, an offline knowledge base construction (\cref{sec:knowledgebase}), online page extraction ((\cref{sec:page_extract_and_analysis}) and knowledge-driven repair phases (\cref{sec:knowledgeRepair}). 

For the offline knowledge base construction phase (\cref{fig:overview}-A), we built a knowledge base (KB) consisting of two parts, i.e., a low-level Component Knowledge Base (KB-Comp) and a high-level System Design Knowledge Base (KB-System). This knowledge base functions as a domain expert, offering guidance for addressing potential UI issues. 
Given the frontend code and rendered page, we employ a parallel dual-pipe method to extract the used components and their corresponding property groups (as shown in \cref{fig:overview}-B).
Finally, we implement a knowledge-driven, LLM-based repair method (\cref{fig:overview}-C). This approach allows us to carefully analyze and repair issues concurrently. By employing a divide and conquer strategy, we tackle each component/property group separately before integrating the repairs. This ensures a cohesive, optimized final output, achieved through a thorough and scrutinized repair process.

\subsection{Knowledge Base Construction}
\label{sec:knowledgebase}

As introduced in \cref{sec:materialdesign}, we utilize Material Design 3 as our foundational knowledge base. 
By systematically analyzing the guidelines for efficient knowledge retrieval, we restructured them into two parts: a low-level Component Knowledge Base (KB-Comp) and a high-level System Design Knowledge Base (KB-System). 
KB-Comp provides detailed guidance for atomic-level elements, while KB-System addresses the broader aspects of aesthetic and functional harmony in interface design. 
For example, a guideline from KB-Comp might ensure a button is sized appropriately for accessibility purpose, while a guideline from KB-System would focus on how this button fits into the overall page layout and flow.
Within these knowledge bases, we further classified the information into soft constraints (i.e., \textit{recommended} guidelines) and hard constraints (i.e., \textit{do/don't} guidelines), catering to different levels of design concerns.
For instance, a soft constraint would suggest font styles that align with the brand's aesthetic, whereas a hard constraint may specify the minimum text size for readability.

\begin{figure}
    \centering
    \label{fig:component-distribution}
    \begin{subfigure}[b]{0.45\textwidth}
        \centering
        \includegraphics[width=\linewidth]{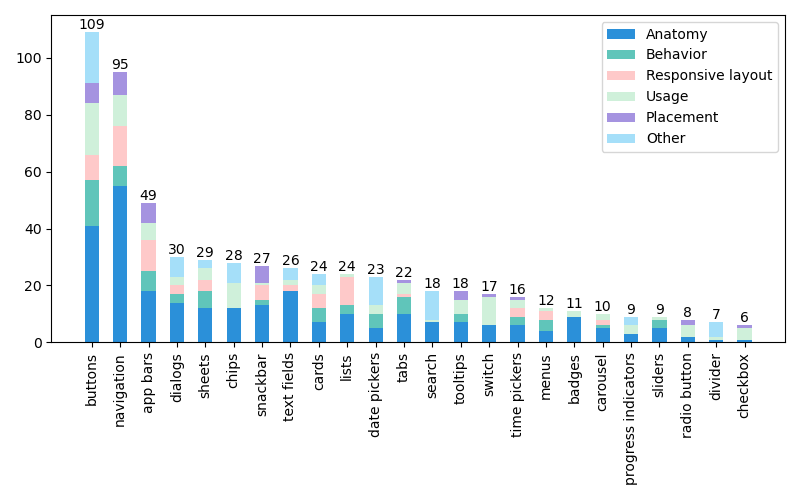}
        \captionsetup{font=scriptsize}  
        \caption{Distribution by Component Type}
        \label{fig:dis1}
    \end{subfigure}
    \hfill 
    \begin{subfigure}[b]{0.45\textwidth}
        \centering
        \includegraphics[width=\linewidth]{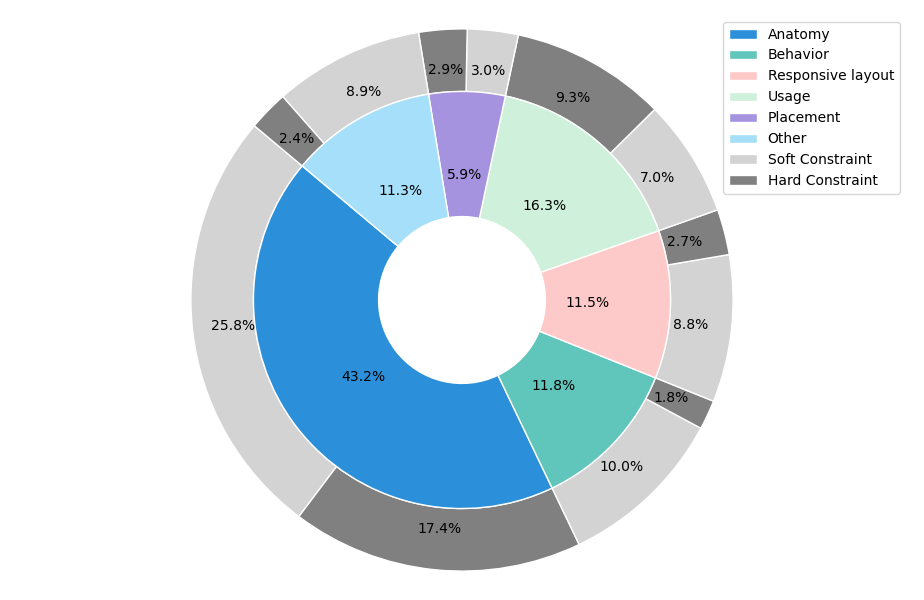}
        \captionsetup{font=scriptsize}  
        \caption{Distribution by Component Design Aspects with Soft/Hard Constraint. 
        }
        \label{fig:dis2}
    \end{subfigure}
    \caption{Distribution of Component Guidelines
    }
    \vspace{-4mm}
\end{figure}

\subsubsection{\textbf{Component knowledge base}}
\label{sec:componentKB}

We created the component knowledge base (KB-Comp) at the component level, drawing from \texttt{Components} section in the Material Design (see \cref{sec:materialdesign}). Each component is linked with a range of guidelines. In total, we obtained 24 component types associating with Material Design guidelines. \cref{fig:dis1} illustrates the number of guidelines for each component type and the distribution of various UI/UX design aspects (like usage, responsive layout) that these guidelines address. 
The distribution of guidelines across these component types demonstrates a notable imbalance, showing a long-tail distribution pattern. For instance, buttons and navigation components are the most abundant, with 109 and 95 entities respectively, while checkboxes are the least represented, with only 6 entities. 
To facilitate understanding, \cref{fig:kbexamples} (a) shows two hard constraints and two soft constraints regarding the button component. In this example, the first hard constraint aligns with the guideline shown in \cref{fig:exp2-hard}, and the first soft constraint correlates with the guideline in \cref{fig:exp2-soft}.

We further conducted an in-depth analysis of our component knowledge base in terms of design aspects, identifying the distribution of soft and hard constraints within each design aspect.
In total, our KB-Comp includes 228 hard constraints (113 \textit{do}s, 115 \textit{don't}s), and 399 soft constraints (\textit{recommended} guidelines).
\cref{fig:dis2} shows the distribution of different design aspects and the percentages of hard and soft constraints. 
The anatomy aspect comprises 43.2\% of the guidelines, whereas placement represents 5.9\%. 
Notably, the number of soft constraints frequently matches or surpasses the number of hard constraints. This suggests that while soft constraints might be less mandatory, they offer a wealth of valuable information to convey better user experience.

In conclusion, based on our detailed knowledge base, we propose knowledge-driven LLM-based method, which can effectively manage the long-tail distribution of design guidelines, ensuring an equitable application of these principles across all component types.

\subsubsection{\textbf{System Design knowledge base}}
\label{sec:GKB}

To mitigate the challenge of hallucinations in LLMs and enhance their analysis capabilities, our approach involves a precise association of high-level design guidelines with actionable properties. This is based on the understanding that properties are the direct targets for addressing high-level design issues.

\begin{figure}
    \centering
    \includegraphics[width=0.9\linewidth]{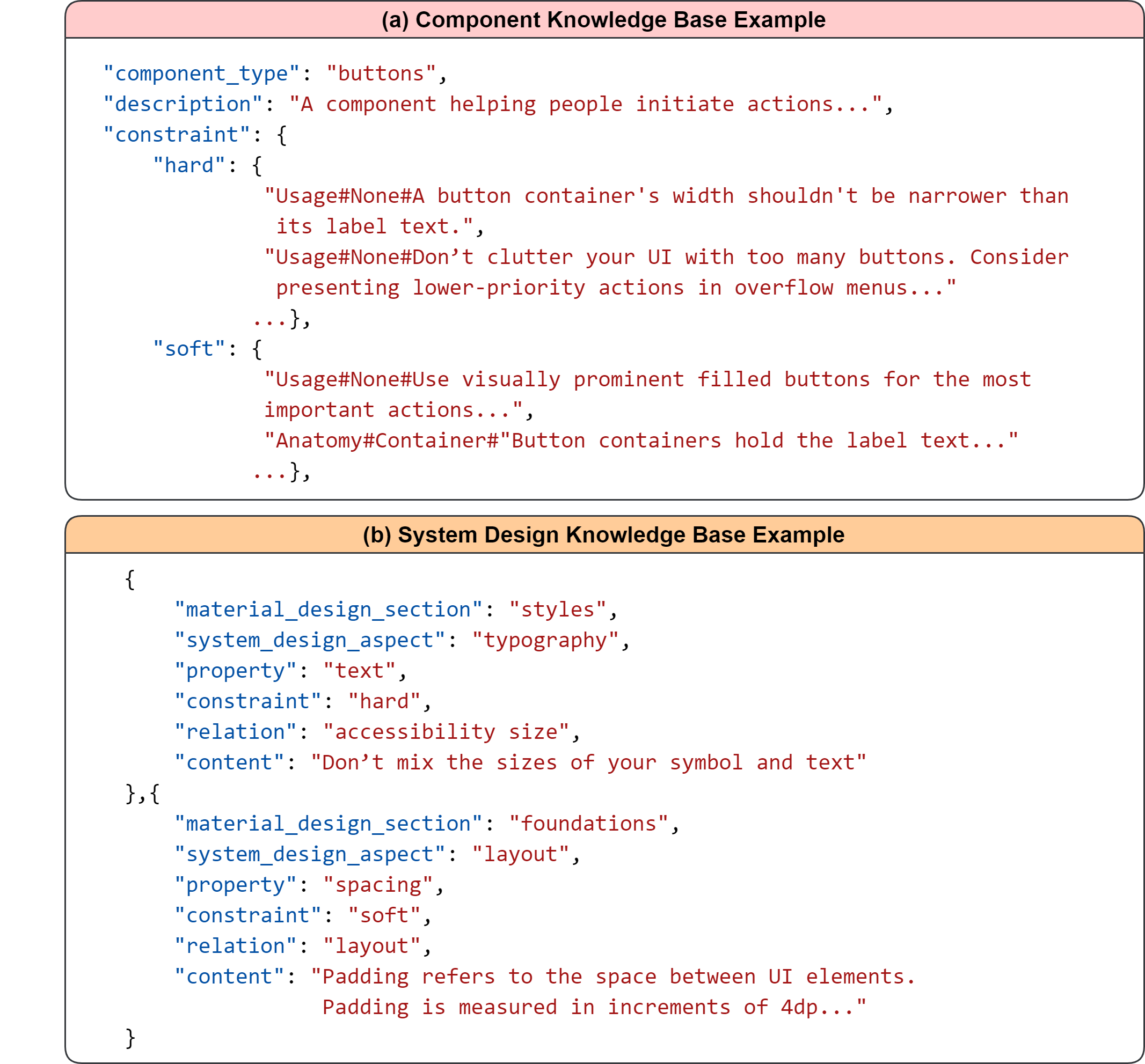}
    \caption{ Examples of Component and System Design Knowledge entities
    }
    \vspace{-6mm}

    \label{fig:kbexamples}
\end{figure}

To this end, we reformulated high-level design guidelines, i.e.,  \texttt{Foundations} and \texttt{Styles} sections (see \cref{sec:materialdesign}), into a structured system design knowledge base (KB-System). In total, We identified 12 system \underline{design aspects} (e.g., typography), summarized 7 \underline{properties} (e.g., clickable and spacing) and 15 \underline{mapping relations} between them. 
\revision{For example, design aspects `\textit{role}' and `\textit{utilities}' in \texttt{Styles} - Color section, as well as the design aspect accessibility `\textit{contrast}' in \texttt{Foundations} - Color section are both connected to the `color' property.} These relations are crucial for marking guidelines, aiding LLMs in understanding the intricate relationships between design elements and their practical applications in a web development context.
Based on this, we reconstructed the design guidelines into triples of \textit{$<$system design aspect, property, relation$>$}. \cref{fig:kbexamples} (b) shows two examples of the system design knowledge base.

\textbf{Foundations}
have 7 key aspects including Flow, Layout, Implementation, Structure, Label, Text, and Color. 
For instance, the `Flow' aspect relates to how users navigate through an interface, while `Color' pertains to both aesthetic appeal and accessibility. In total, we identified 118 knowledge entities for Foundations.

\textbf{Styles}
consist of 5 key aspects: Typography, Elevation, Icons, Color, and Shape. These elements craft the UI’s visual identity, adaptable through Material themes. In our work, we structured 80 knowledge entities related to Styles, with Icons specifically linked to icon buttons and Shapes applicable across all component types.

\textbf{Property Groups} 
include 7 categories: Group, Clickable, Spacing, Platform, Label, Text, and Color. These are actionable property levels that we map to high-level guidelines for effective detection and repair.

\textit{(1) Group}: This property is associated with the layout and accessibility elements within the UI, focusing on aspects such as landmark recognition and prioritization in navigation. It is linked to the `Structure', `Flow', and `Layout' of Foundations.

\textit{(2) Clickable}: This property predominantly relates to interactive elements like buttons, focusing on accessibility aspects such as focus order and target size. It is also connected to the 'Elevation' style, emphasizing floating elements.

\textit{(3) Spacing}: Integral to layout design, this property focuses on the spatial arrangement and coherence of UI elements. It is linked to the `Layout' aspect in Foundations.

\textit{(4) Platform}: This property addresses the layout implementation across different platforms, ensuring accessibility and consistency in various environments. It is associated with `Implement' and `Layout' in Foundations.

\textit{(5) Label}: Expanded to encompass alternative text and captions for images, videos, icons, and buttons, this property plays a pivotal role in enhancing accessibility. It is linked to `Label' for content accessibility in Foundations, as well as `Structure' for accessibility labeling.

\textit{(6) Text}: This property group covers aspects of typography and content accessibility size. It is connected to the `Typography' style and the `Text' and `Structure' aspects of Foundations, focusing on size and accessibility in headings.

\textit{(7) Color}: Color is intricately linked to both the aesthetic and functional roles in UI design, including accessibility considerations like contrast. This property group is connected to the `Color' aspect in both Styles and Foundations.

\subsection{Page Extraction}
\label{sec:page_extract_and_analysis}
In this phase, given the code and the rendered page as inputs, \tool{} examine the usage of component in the code, and system design aspects in the rendered output of the page. 

This dual analysis is important as it bridges the gap between technical code and the user perceived experience. It allows for detailed exploration of component usage within the code, alongside addressing broader design concerns evident in the rendered output at the property group level.

\subsubsection{\textbf{Components Extraction}}
The first aspect of our strategy is to identify a list of component types within the Page Under Test (PUT) using GPT-4. This initial step is important for the efficient retrieval of relevant component knowledge from KB-Comp in subsequent repair stages. 

Given a PUT code \( In_{\text{PUT}} \), we propose a systematic component extraction prompt \( P_{\text{comp\_extra}} \) to analyse a PUT, represented by the equation:
\vspace{-1mm}
\[ B1 = P_{\text{comp\_extra}}\left( In_{\text{PUT}} \right) \]

This process results in a list of component type names from the PUT, defined by:
\( B1 = \{ \text{Component}_i \,|\, i = 0, 1, \ldots, k \} \)

\subsubsection{\textbf{Property Extraction}}
The second aspect of our extraction strategy leverages the Playwright tool~\cite{microsoft_playwright_2024}, a robust testing tool developed by Microsoft, to dissect the rendered frontend page into various properties. This approach is critical for understanding the complex relationships between different page elements and their various properties, which are integral to addressing broader design issues.

Properties are analyzed to ensure consistency in design elements such as spacing and color schemes, as well as overall layout coherence. This stage focuses on elements that are more readily apparent in the rendered output rather than just within the code itself. The use of the Playwright tool enables us to pre-render the page and extract key DOM information such as XPath, color, framed within seven specified categories: Group, Clickable, Spacing, Platform, Label, Text, and Color.

We process these properties to produce \textit{$<$outerHTML, property$>$} pairs, yielding code snippets and property values. This meticulous organization of data is crucial in pinpointing problem areas within the code, thereby facilitating an in-depth and thorough analysis by the LLM. It enhances the efficacy of the subsequent repair phase. The output from this stage is represented as \( B2 \), a list of 7 extracted property groups.

\subsection{Knowledge-Driven Repair}
\label{sec:knowledgeRepair}

In the knowledge-driven repair phase, we leverage the reasoning capabilities of LLMs coupled with a broader spectrum of knowledge bases from \cref{sec:knowledgebase} to perform detailed repairs at both the component and property levels of frontend code. This phase marks a critical transition from the initial analysis to the actual refinement and enhancement of the code based on the insights gained. 

The process is divided into three key steps.
For each component or property group, we first retrieve the related guidelines from our knowledge bases (\cref{fig:overview} Step C1). 
With related knowledge/guidelines and associated component/property group, we adopt LLMs to conduct targeted repair (\cref{fig:overview} Step C2).
This method ensures that the specific needs of various components and properties are addressed with focused attention.
Finally, we assemble, reassess and synthesize the individual repairs into a cohesive, optimized, and final repaired output (\cref{fig:overview} Step C3).
By adopting the divide-and-conquer strategy, we ensure detailed, thorough, and holistic repair.

\subsubsection{\textbf{Relevant Knowledge Retrieval}}
In this step, we first preprocess the retrieved components to map them with our component knowledge base, and then retrieve corresponding knowledge from KB-Comp and KB-System.

\textit{(a) Component \revision{Mapping and} Knowledge Retrieval:}
To map extracted web page components to the component names in our KB-Comp for alignment, we use LLMs for reasoning with the prompt \( P_{\text{map\_kb}} \). This approach allows us to handle the diversity in terminology used across different front-end libraries such as Shadcn\cite{shadcn_ui_2024}, Tailwind CSS\cite{tailwindcss_2024} without the need of defining \revision{any exhaustive rule-based} mapping function.
For instance, a `navigation bar' component might be referred to as `Navbars' in Tailwind CSS, while in Shadcn UI it could be labeled as `Navigation Menu'. Such variability in naming conventions can be effectively addressed by LLMs without the need to define an exhaustive mapping function. 
The mapping process is represented by:
\vspace{-1mm}
\[ B1^* = P_{\text{map\_kb}}\left( B1_i \,|\, i = 0, 1, \ldots, k \right) \]
\vspace{-1mm}
Where \(B1^*\) is the resultant list of components mapped to our KB-Comp equivalents. The corresponding knowledge items for \(B1^*\) are then retrieved from KB-Comp, denoted as \(C2_{\text{comp}}\).

\textit{(b) Property Knowledge Retrieval:}
For each property group in \(B2\), we directly retrieve related knowledge items from our KB-System, denoted as \(C2_{\text{property}}\).

\textit{(c) Aggregation:}
Finally, we aggregate the extracted information and the retrieved knowledge as the following:
\[ C1 = \{ B1^* \cup B2_i \,|\, i = 0, 1, \ldots, j \} \]
\[ C2 = \{ C2_{\text{comp}} \cup C2_{\text{property}_i} \,|\, i = 0, 1, \ldots, h \} \]

In \(C1\), we have a list of mapped components and property types extracted from the PUT. Furthermore, \(C2\) contains the corresponding knowledge entities from our knowledge bases. Examples of these entities are illustrated in Fig \ref{fig:kbexamples}, serving as preparation for the repair stages.

\subsubsection{\textbf{Individual Repair}}
In the individual repair phase, we implement an LLM-driven, detail-oriented process to inspect and repair frontend code in relation to specific components and property groups. This phase also employs a distinct approach for handling soft and hard constraints to guide the analysis and repair process (denoted as $C3$).
The output of this phase is formatted as a JSON object, comprising sections for poor design code snippets, associated guideline references, and repair suggestions.

Formally, given a PUT code \( In_{\text{PUT}} \) alongside the extracted lists \( C1 \) and their corresponding knowledge lists \( C2 \), we apply an iterative repair prompt \( P_{\text{individual}} \):
\vspace{-1mm}
\[ C3 = P_{\text{individual}}\left( In_{\text{PUT}} \,||\, C1 \,||\, (C2_{\text{soft}} \,||\, C2_{\text{hard}}) \right) \]
\vspace{-1mm}
In the LLM prompting stage, we differentiate the guiding narrative for soft and hard constraints. For hard constraints (denoted as \( C2_{\text{hard}} \)), we instruct the LLM as follows:  \textit{``Here are the guidelines you must follow, we name it `hard constraints'. 
Remember this is mandatory. Once you find a bad design not following the guideline, you must fix it.''}
This directive emphasizes the obligatory nature of these constraints in the repair process.

Conversely, for soft constraints (denoted as \( C2_{\text{soft}} \)), we prompt the LLM with: \textit{``Here are the general guidelines you can use, we name it `soft constraints'. 
Remember this is not mandatory, regarded as optional.''}
This approach suggests a more flexible consideration of these guidelines, allowing for discretionary application in the repair process.

The resulting \( C3 \) list comprises the code in question, the associated knowledge, and repaired code snippets for each component and property:
\[ 
C3 = \left\{ 
\begin{array}{l}
\textit{``bad\_design\_code''}_i, \\
\textit{``detailed\_reference\_guidelines''}_i, \\
\textit{``suggestion\_fix\_code''}_i
\end{array}
\,|\, i = 0, 1, \ldots, m 
\right\} 
\]

This structured approach enables the LLM to deliver nuanced repair suggestions, tailored to the specific context and requirements of each detected issue, aligning closely with the established design principles and usability guidelines.

\subsubsection{\textbf{Overall Repair}}
In this final step, we integrate the repair suggestions obtained from both the component-level and property-group analyses. This integration ensures a holistic repair of the frontend code.

Formally, given the original PUT code \( In_{\text{PUT}} \) and the set of partially repaired code snippets \( C3 \), our objective is to use these inputs to produce the final, fully repaired page. To achieve this, we employ prompt with \( P_{\text{all}} \), which is designed to direct the overall repair process:
\vspace{-1mm}
\[ \text{Repaired\_page} = P_{\text{all}}\left( In_{\text{PUT}} \,||\, C3 \right) \]
\vspace{-1mm}
The \( P_{\text{all}} \) prompt effectively merges the various repair suggestions, taking into account any potential overlaps or conflicts between the proposed fixes. The goal is to ensure that the final output, represented as \( \text{Repaired\_page} \), is not only free from identified design flaws but also aligns with the established design guidelines and usability standards. \revision{Due to the page limit, all prompts can be obtained at our GitHub repository.
}

\section{Experiments}
\label{sec:experiments}

In this section, we evaluate the effectiveness and usefulness of our proposed \tool{} by investing three research questions (RQs):

\begin{itemize}
    \item \revision{RQ1: How effective is our approach in detecting violations according to Design Guidelines compared to the baseline?}
    \item RQ2: How effective are individual strategies and the overall repair approach?
    \item RQ3: What is the perceived quality of the fixes generated by our approach from user perspectives?
\end{itemize}

\subsection{Subjects}
\label{sec:subjects}

\revision{
We conducted two experiments to evaluate the performance of our frontend code repair approach on both generated code from Vercel's V0 and real projects from GitHub.
}

\subsubsection{\revision{Vercel’s V0 Projects}}
\label{sec:vercel_dataset_labeling}

\revision{
Our research focuses on examining code and the rendered pages generated by LLM tools, specifically targeting adherence to design guidelines and enabling code-level adjustments. Our work aims to assist developers with limited knowledge of UI design guidelines who utilize LLM-generated UI mock-ups as starting points. Vercel’s V0~\cite{v0dev2024} generated projects were chosen for this purpose, as it is a leading LLM tool that attracted 100K users within three weeks of its release~\cite{v0report}. Additionally, Vercel's V0 is widely used to assist in frontend development, with 259K results on GitHub as of December 2024. }

Vercel’s V0 platform enables users to experiment with and generate UI designs based on their provided prompts. By default, users' attempts are shared publicly, allowing for community visibility. Additionally, Vercel’s V0 manually reviews prototypes created with their tool, showcasing those that are notably good and representative.

We randomly chose 64 projects from Vercel's V0 featured projects and user submissions to ensure a diverse selection. Each project was fully examined based on the following rules:

\revision{(1) Projects with specific and clear page definitions were included, while those with vague prompts like ``give me an icon'' without additional explanation were excluded.}
\revision{(2) Projects included different use cases such as landing pages, product pages, and login pages.}
\revision{(3) Projects that were well-designed and had undergone at least three iterations of human modification and prompting were included.}

Consequently, we selected 20 high-quality human-AI collaborative front-end code projects generated by Vercel's V0. 

\revision{To conduct recognition of guideline violations, two of the authors first discussed and manually reviewed the guidelines and examples on the Material Design guidelines to reach a common understanding. They then independently reviewed the selected projects, manually identifying components, property groups, associated design guidelines, and design violations. In cases of disagreement, a third author was consulted to review the relevant guidelines and provide their judgment, which was then referred to for resolution.}

Through careful discussion and consensus on the annotations, we pinpointed 669 components/property groups across these projects, corresponding to 3,765 design guidelines, of which 196 were violated. 
Note that on average, each project is associated with 188.25 design guidelines. This statistic underscores the complexity and the extensive range of design principles that need to be considered in modern UI development.

\subsubsection{\revision{Github Projects}}

\revision{To further evaluate the effectiveness of our method in real-world projects, we searched for ``top frontend projects'' and ``frontend example projects'' on GitHub and Google, identifying the top 6 popular projects. These projects represent various real-world website uses, ranging from commercial product frontend websites to blogs, with star counts ranging from 1.1k to 5.6k. 
Their popularity and representativeness have proven their usefulness and quality.
}

\revision{We randomly selected 66 files out of 577 files from these projects, ensuring diverse functionalities. These projects utilize a variety of representative packages, including React~\cite{react_dev_2024}, Next.js~\cite{nextjs}, Material-UI~\cite{materialui}, and TypeScript~\cite{typescript}, which can examine the capabilities of our \tool{} in detecting and repairing design issues across these different technologies. 
}
\revision{Following the same protocol in Section~\ref{sec:vercel_dataset_labeling}, we identified 1,793 components/property groups across these projects, of which 115 guidelines were violated.}

\subsubsection{\revision{Subject Analysis}}
\label{sec:subjects_analysis}

\revision{For Vercel's V0 projects, there were 33.45 components/property group per file and 9.8 guideline violations per project. Of these, 149 (76\%) were component-related and 47 (24\%) were property-related. Additionally, 27 (14\%) were soft constraint violations, while 169 (86\%) were hard constraint violations.}
\revision{For GitHub projects, there were 27 components/property group per file and 1.74 guideline violations per project. Among these, 60 (52\%) were component-related and 55 (48\%) were property-related. Soft constraint violations accounted for 15 (13\%), and hard constraint violations accounted for 100 (87\%). }

\revision{Based on our analysis, the generated frontend code from Vercel's V0 was found to have complexity comparable to real GitHub projects (33.45 vs. 27 components/property group per file). This suggests the potential of LLM-generated code to match the structural richness required for real frontend development.}
\revision{
However, GitHub projects have significantly fewer violations per project compared to Vercel's V0 projects (1.74 vs. 9.8), highlighting a substantial gap in quality. This indicates that while LLMs can generate complex code, their outputs are prone to a higher rate of errors and require enhanced automated quality assurance mechanisms to improve code quality.
}
\revision{In addition, Vercel's V0 projects have a higher percentage of component-related issues (over 52\%) compared to property-level issues. Real-world projects, although generally better designed, still have minor issues such as inconsistent padding or missing alt-text.}


\revision{
In conclusion, we found that human-generated issues differ significantly from AI-generated ones, making it valuable to analyze them separately for better insights.}

\subsection{\revision{RQ1: Effectiveness in detecting violations}}
\label{sec:rq1}

\subsubsection{\textbf{Experimental Setup}}
\label{sec:rq1_protocol}

\revision{We used UIS-Hunter~\cite{yang2021don} as our baseline, as it also focuses on detecting design violations against the material design. 
It primarily focuses on do/don't guidelines (i.e., hard constraints), and only supports 23 types of UI components and examines 126 visual design guidelines.
In comparison, our tool collects 399 guidelines for 24 UI component types, including both do/don't and general guidelines. 
}

\revision{To make a fair comparison, we selected corresponding instances involving hard constraint violations: 127 issues from Vercel's V0 and 48 issues from GitHub, to perform the experiment. We adopted the precision and recall metrics. 
A detection result is considered as a true positive if the elements are actually faulty.
}

\revision{
During the experiment, we identified limitations in UIS-Hunter due to its rule-based approach, which struggles with component mapping. For instance, elements like buttons within a navigation bar should be classified under multiple types, but UIS-Hunter restricts them to a single type. Moreover, UIS-Hunter was specifically designed for Android apps and lacks the capability to handle the complexity of web components commonly used in frameworks such as React and Tailwind CSS.
}
\revision{
To address these issues and adapt UIS-Hunter for the experimental data (as some component names in the experimental data differ from those supported in UIS-Hunter), we focused on evaluating its performance in detecting guideline violations. 
Two authors manually mapped component names to their corresponding types, which took about one hour per project. In contrast, our method enables automated component analysis and mapping, demonstrating significantly greater flexibility than the rule-based approach.}

\subsubsection{\textbf{Results}}

\revision{
As shown in \cref{tab:baselinecompare}, for recall, UIS-Hunter achieves 32.3\% for Vercel's V0 projects and 22.9\% for GitHub projects, while our method achieves 92.1\% and 83.3\%, respectively.}

\revision{
UIS-Hunter's lower true positive rates stem from its reliance on fixed rules and thresholds, which result in many issues being overlooked.
We identified three primary failure reasons for UIS-Hunter.
First, it is unable to evaluate active state components. 
It cannot assess the correctness of components in active states and can only identify issues in static screenshots through image processing. 
Second, UIS-Hunter fails to detect responsive design issues such as line breaks not displaying correctly in larger windows.
Addressing these issues would require multiple detections, a capability that this method lacks.
Moreover, it cannot capture the semantics of components, resulting in an inability to detect subtle component issues. For instance, UIS-Hunter cannot distinguish and evaluate the semantic differences between buttons for ``follow'' and ``message'', which require different active colors based on their semantics.
}

\revision{In contrast, our algorithm overcomes these limitations, ensuring high recalls and consistency detection among components, such as ensuring uniform font colors across multiple buttons. It also accurately identifies minor visual violations, such as padding differences between the left and right, which purely visual-based methods may miss.}

\revision{For precision, as shown in \cref{tab:baselinecompare}, UIS-Hunter achieves 55.4\% for Vercel's V0 projects and 47.8\% for GitHub projects, while our method achieves 87.3\% and 90.9\%, respectively.}
\revision{UIS-Hunter's lower precision is largely due to false positives caused by OCR errors, such as spaces in long text strings. 
Additionally, its rigid rule-based approach-such as requiring colored dividers in lists or limiting the number of icons-further contributes to the higher false positive rate.}

\revision{Overall, our comparative analysis with the baseline demonstrates that our method effectively addresses component guideline mapping issues caused by different language packages, accurately detects guideline violations in LLM-generated code, and is extendable to real-world projects.}

\subsection{RQ2: Effectiveness in repairing violations}
\label{sec:rq2}

\subsubsection{\textbf{Experimental Setup}}
\label{sec:rq2_protocol}

\revision{
To evaluate the effectiveness of \tool{} in correcting guideline violations in frontend code, we conducted an assessment of each step module. This evaluation includes analyzing Component Extraction, Component Mapping, Violation Detection, various aspects of Violation Repair, and the overall performance of the tool.
}

\revision{In the evaluation, we compared outputs with manually annotated ground truth for each step. For "Component Extraction", an extraction is deemed correct if the component name can be found in the file. For "Component Mapping", a mapping is considered as correct if the mapped type matches an entry in the Material Design component list. Successful detection and repair are defined as correctly identifying and fixing faulty elements based on the guidelines. 
}

\revision{To evaluate the effectiveness of component/property and soft/hard constraint strategies separately, we tested each strategy in isolation and evaluate them on their respective guideline subsets.
For example, for component-only violation repair strategy, we disabled the property repair module and calculated results only for component-specific guidelines.}

\begin{table}[ht]
\vspace{-2mm}
\centering
\caption{\revision{Comparison of Violation Detection Performance}}
\resizebox{0.8\columnwidth}{!}{%
\begin{tabular}{lcccc}
\toprule
 & \multicolumn{2}{c}{\textbf{Vercel's V0 Projects}} & \multicolumn{2}{c}{\textbf{GitHub Projects}} \\
\cmidrule(lr){2-3} \cmidrule(lr){4-5}
 & \textbf{Rec.} & \textbf{Prec.} & \textbf{Rec.} & \textbf{Prec.}  \\
\midrule
UIS-Hunter & 32.3 & 55.4 & 22.9 & 47.8 \\
\tool{} & 92.1 & 87.3 & 83.3 & 90.9  \\
\bottomrule
\end{tabular}
}
\label{tab:baselinecompare}
\vspace{-2mm}
\end{table}

\begin{table}[ht]
\vspace{-2mm}
\centering
\caption{\revision{Performance Metrics for Each Key Strategy in Vercel's V0 and GitHub Projects}}
\resizebox{0.9\columnwidth}{!}{%
\begin{tabular}{lcccc}
\toprule
 & \multicolumn{2}{c}{\textbf{Vercel's V0 Projects}} & \multicolumn{2}{c}{\textbf{GitHub Projects}} \\
\cmidrule(lr){2-3} \cmidrule(lr){4-5}
 & \textbf{Rec.} & \textbf{Prec.} & \textbf{Rec.} & \textbf{Prec.} \\
\midrule
Component Extraction & 96.7 & 93.6 & 96.5 & 91.8 \\
Component Mapping & 94.1 & 90.5 & 94.9 & 90.2 \\
Violation Detection & 92.9 & 90.1 & 87.8 & 93.5 \\
\midrule
Violation Repair (Comp.) & 89.9 & 87.0 & 81.7 & 87.5 \\
Violation Repair (Property) & 87.2 & 85.4 & 89.0 & 94.2 \\
\midrule
Violation Repair (Soft) & 70.4 & 82.6 & 67.7 & 77.0 \\
Violation Repair (Hard) & 92.3 & 87.2 & 88.0 & 92.6 \\
\midrule
\textbf{Violation Repair (All)} & \textbf{89.3} & \textbf{86.6} & \textbf{85.2} & \textbf{90.7} \\
\bottomrule
\end{tabular}
}
\label{tab:substepcompare}
\vspace{-2mm}
\end{table}

\subsubsection{\textbf{Results}}

\revision{\textbf{Component Extraction Results:}}
\revision{Upon examining \tool{}'s performance in Component Extraction, we identified 88 component types across 20 Vercel's V0 projects and 64 component types for GitHub projects. As shown in \ref{tab:substepcompare}, \tool{} achieved 96.7\% recall and 93.6\% precision on Vercel's V0 projects , and 96.5\% recall and 91.8\% precision on GitHub projects. Precisions are slightly lower than recalls because the LLM used an aggressive strategy to extract more components when uncertain, minimizing omissions. 
We found that the missed components in our detection had little impact on mapping performance because they were usually the same type as the correctly identified components, such as ``buttons'', differing only in their specific names, like ``likebutton'' or ``sharebutton''.
}

\revision{\textbf{Component Mapping Results:}}
\revision{We identified 88 component types in 51 Vercel's V0 projects and 39 component types in GitHub projects. As shown in \ref{tab:substepcompare}, our model achieved 94.1\% recall and 90.5\% precision on Vercel's V0 projects, and 94.9\% recall and 90.2\% precision on GitHub projects.}

\revision{Complex components formed from simpler ones, such as navigation bars, cards and lists, achieved nearly 100\% precision and recall. The LLM effectively analyzed and identified these groups, which is challenging for rule-based methods. However, simple components defined primarily by CSS, like badges and dividers, were less effectively extracted by our method. }

\revision{In general, the results demonstrate that the LLM-based analysis method can handle complex component extraction tasks and can deal with various, sometimes anecdotal implementation behaviors, such as using hyperlinks instead of proper button tags to represent a button.
Rule-based approaches struggle to correctly identify such variations.}

\revision{\textbf{Violation Detection Results:}}
\revision{We evaluated guideline violations on 196 Vercel's V0 and 115 GitHub's. As shown in \cref{tab:substepcompare}, \tool{} achieved 92.9\% recall and 90.1\% precision on Vercel's V0 projects, and 87.8\% recall and 93.5\% precision on GitHub projects. Our method effectively detects both component and property issues. For components, it identifies violations in anatomy, behavior, layout, usage, and placement. For properties, it detects issues related to labels, text, color, and spacing, focusing on accessibility problems like alt-text and labeling.
}


In our analysis, our method demonstrates higher recall of candidate fixes on Vercel’s V0 compared to GitHub projects. However, recall decreases on complex GitHub pages as \tool’s LLM avoids significant adjustments to prevent introducing new issues.

\revision{\textbf{Component/Property Only Repair Results:}}
\revision{For component-only repair, we evaluated 149 Vercel's V0 violations instances and 60 GitHub's. As shown in \cref{tab:substepcompare}, our method achieved 89.9\% recall and 87\% precision on Vercel's V0, and 81.7\% recall and 87.5\% precision on GitHub projects. For property-only repair, we evaluated 47 Vercel's V0 violations instances and 55 GitHub's. Our model achieved 87.2\% recall and 85.4\% precision on Vercel's V0, and 89\% recall and 94.2\% precision on GitHub projects.}

\revision{Our \tool{} demonstrated better performance in component repairs for Vercel's V0 projects, while property repairs proved more effective for GitHub projects. Upon analysis, we discovered that GitHub projects often contained minor accessibility issues, such as missing alt-text, improper labeling, or errors caused by handwriting, likely due to less rigorous developer writing and review processes. These types of issues are generally easier for LLMs to predict and repair. In contrast, LLM-generated frontend projects present a challenge for \tool{} to predict values accurately, even with reference sizes from rendered pages. LLMs tend to predict larger values to ensure accessibility, which can compromise page consistency.}

\revision{\textbf{Soft/Hard Constraint Only Repair Results: }}
\revision{For soft constraint-only repair, we evaluated 27 Vercel's V0 violations instances and 15 GitHub's. Our model achieved 70.4\% recall and 82.6\% precision on Vercel's V0 projects, and 67.7\% recall and 77\% precision on GitHub projects. For hard constraint-only repair, we evaluated 169 Vercel's V0 projects and 100 GitHub projects. The results show  92.3\% recall and 87.2\% precision for Vercel's V0, and 88\% recall and 90.7\% precision for GitHub projects.}

\revision{The performance of soft constraint repair is generally lower than that of hard constraint repair.
On GitHub project, results for hard constraint repairs have lower recall but improved precision. These are due to the conservative approach of LLMs to avoid introducing new issues in complex pages, resulting in fewer true positives and false negatives. On Vercel's V0 projects, soft constraint repairs perform better, showcasing the LLM’s ability to enhance design diversity and aesthetics when the tested pages have fewer hard constraint issues.}

\begin{figure}
    \centering

    \begin{subfigure}[b]{0.23\textwidth}
        \centering
        \includegraphics[width=\textwidth]{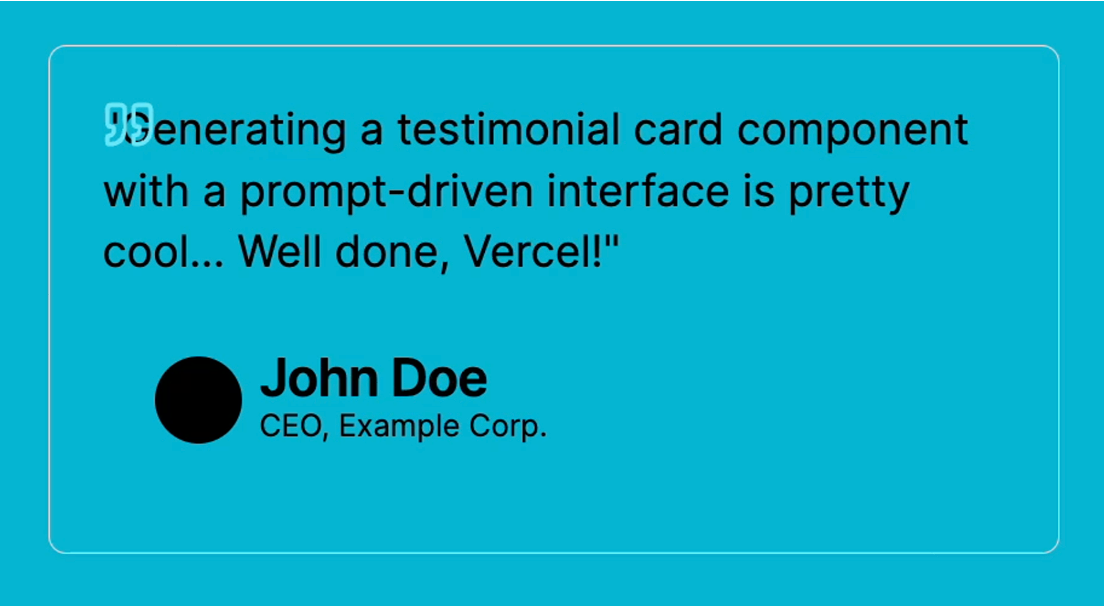}
        \captionsetup{font=scriptsize}
        \caption{Example 1 before-repair: demonstrates issues with element overlapping and color contrast.}
        \label{fig:a}
    \end{subfigure}
    \hfill 
    \begin{subfigure}[b]{0.23\textwidth}
        \centering
        \includegraphics[width=\textwidth]{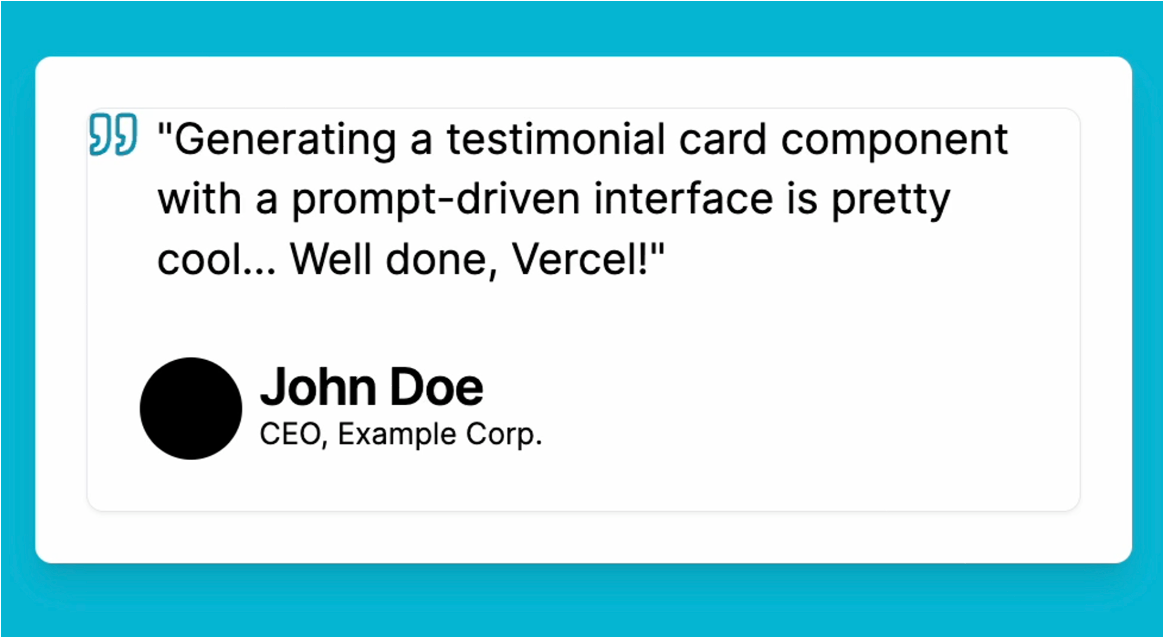}
        \captionsetup{font=scriptsize}
        \caption{Example 1 after-repair: issues resolved, for enhanced accessibility and aesthetics.}
        \label{fig:b}
    \end{subfigure}

    \begin{subfigure}[b]{0.23\textwidth}
        \centering
        \includegraphics[width=\textwidth]{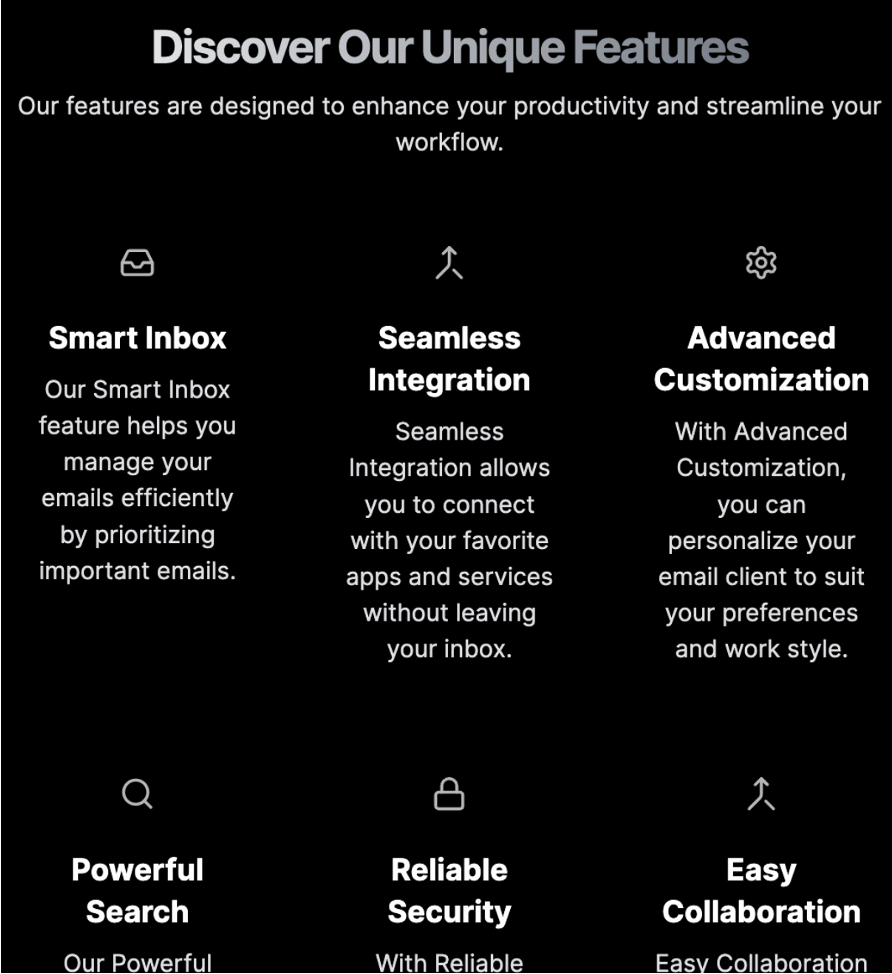}
        \captionsetup{font=scriptsize}
        \caption{Example 2 before-repair: showcases responsive layout challenges in overall design.}
        \label{fig:c}
    \end{subfigure}
    \hfill
    \begin{subfigure}[b]{0.23\textwidth}
        \centering
        \includegraphics[width=\textwidth]{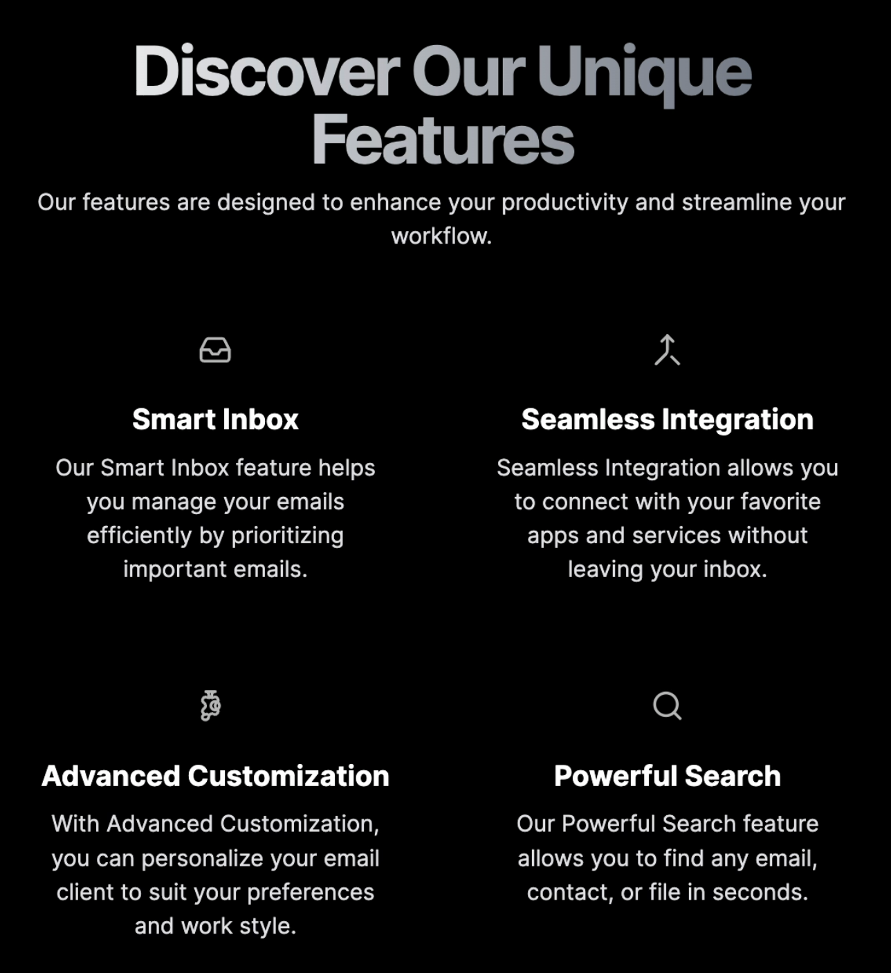}
        \captionsetup{font=scriptsize}
        \caption{Example 2 after-repair: issues resolved, addressing responsive design issues effectively.}
        \label{fig:d}
    \end{subfigure}

    \caption{Examples Illustrating \tool{}'s Repair Effectiveness: Comparative Analysis Before and After Repairs}
    \label{fig:2x2grid}
    \vspace{-6mm}
\end{figure}

\revision{\textbf{Overall Repair Results:}}
\revision{For overall repair, the proposed method achieved 89.3\% recall and 86.6\% precision on Vercel's V0 projects, and 85.2\% recall and 90.7\% precision on GitHub projects.}
Our approach effectively adheres to design guidelines without extensive training. For instance, as illustrated in \cref{fig:a} and \cref{fig:b}, \tool{} adeptly solves element overlapping issue and enhances background color contrast, making the design accessibility-friendly and visually appealing. Furthermore, as showcased in \cref{fig:c} and \cref{fig:d}, our tool effectively addresses responsive layout issues, ensuring optimal appearance across various window sizes. Another striking example, presented in \cref{fig:exp2}, demonstrates our software's proficiency in unifying component colors and improving aspects like accessibility, text legibility, and contrast.

In conclusion, \tool{} demonstrates a remarkable capacity for frontend code repair, significantly enhancing user experience and interface quality. Its holistic approach, blending various strategies and constraints, proves crucial in navigating and rectifying a wide range of design issues, thus establishing it as an effective tool in contemporary frontend development.

\subsection{RQ3: Perceived usefulness}
\subsubsection{\textbf{Procedure}}

A user study was conducted to evaluate the usefulness and perceived quality of the proposed approach.

\revision{We recruited a total of 26 participants (12 female,14 male), aged between 25 and 35. The group comprised 6 researchers, 10 students, and 10 software developers from diverse countries including Australia 8, China 15, and the United States 3.
The educational backgrounds of the participants were varied: 5 had Bachelor’s degrees, 15 had Master’s degrees, and 6 had Ph.D.s. Their affiliations were diverse, with 6 from science agencies, 10 from major internet companies, and 10 from universities. }

To align with our research background, the study used both the original and repaired versions of 20 different projects from \revision{Vercel V0  (as discussed in Section \ref{sec:subjects})}.
\revision{We created two surveys, each containing 10 different pairs of interactive rendered UI page videos, with each participant randomly assigned to one survey. Thus, 13 evaluations per survey and video. Each video pair featured a before and after repair view, and we randomized their order to obscure which was produced by our tool. The videos, each 10 seconds long, displayed the rendered UI page of the code, including interactions with elements and page resizing to demonstrate UI changes and UX. Participants could repeatedly view the videos without time constraints while responding to questions, ensuring detailed observation. }

\revision{For each pair of videos, participants were asked to review and answer the question: ``Please rate each page's visual attractiveness using a 5-point Likert scale.'' After participants completed their ratings, we conducted interviews to reveal which pages were repaired and discuss their scoring rationale. This approach uncovered patterns, such as why some pages were rated higher before repairs. }

\subsubsection{\textbf{Results}}

\begin{figure}[ht]
    \centering
    \includegraphics[width=1\linewidth]{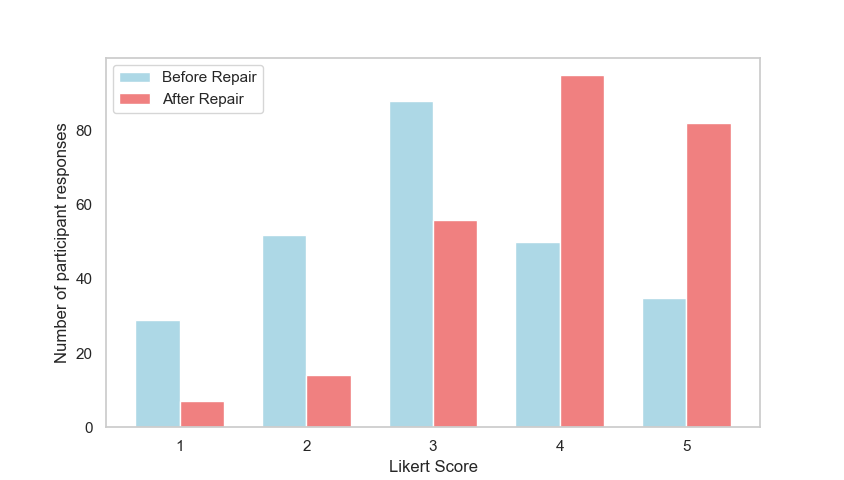}
    \caption{Frequency of Likert Scores Before and After Repair
    }
    \vspace{-6mm}
    \label{fig:u1}
\end{figure}

The user study's results, depicted in Fig. \ref{fig:u1}, illustrate the satisfaction ratings for the 20 subjects' UIs before-after repaired. The Likert scores, ranging from 1 to 5, are distributed along the x-axis with before repair scores in blue and after repair scores in red. Notably, there is a significant shift in participant responses towards the higher scores (4 and 5) after repair, indicating improved user satisfaction. The most frequent before-repair score was 3, shifting to 4 after-repair, clearly signifying enhanced satisfaction with the repaired interfaces.

\begin{figure}[ht]
    \centering
    \begin{subfigure}[b]{0.23\textwidth}
        \centering
        \includegraphics[width=\linewidth]{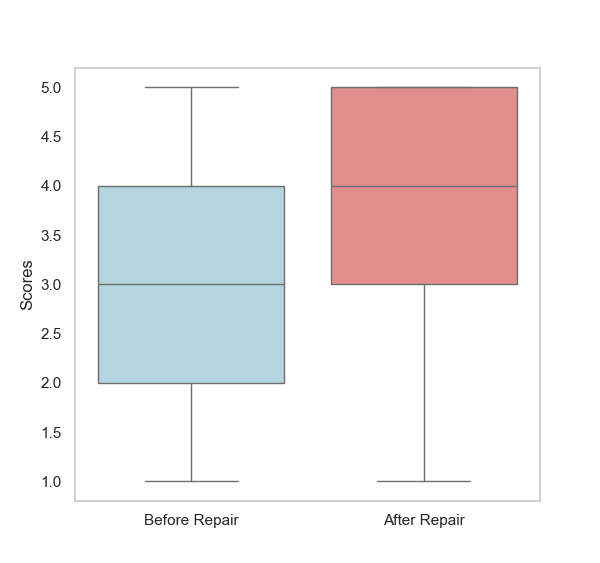}
        \captionsetup{font=scriptsize}
        \caption{Comparison of Scores Before and After Repair}
        \label{fig:u2}
        
    \end{subfigure}
    \begin{subfigure}[b]{0.23\textwidth}
        \centering
        \includegraphics[width=\linewidth]{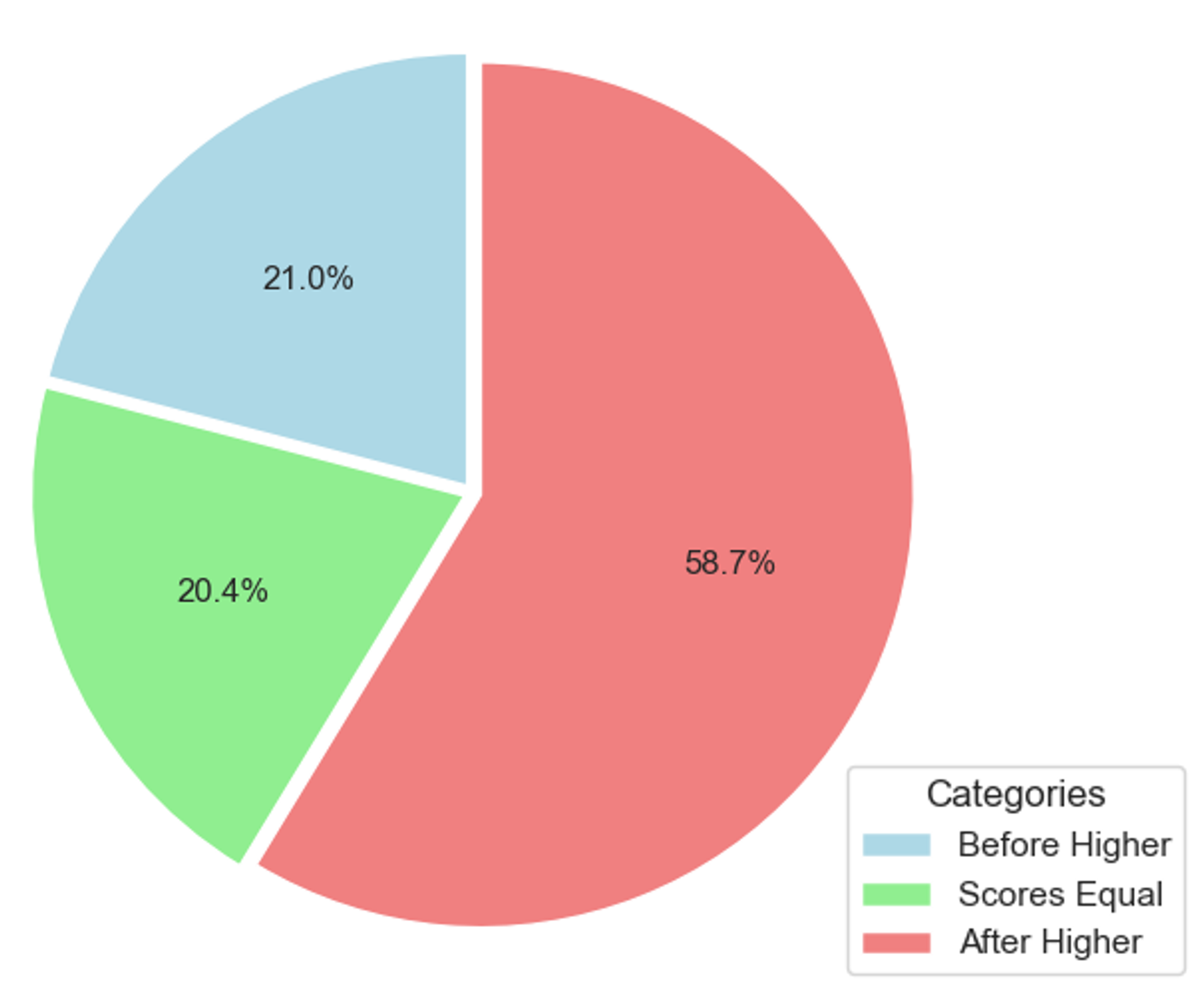}
        \captionsetup{font=scriptsize}
        \caption{Score Comparisons: Before vs After Repair}
        \label{fig:u3}
    \end{subfigure}
    \caption{Comparative Analysis of User Satisfaction}
    \vspace{-4mm}
\end{figure}

For a more rigorous analysis, we employed the Wilcoxon Signed-Rank test. The test yielded a p-value of approximately \( p \approx 6.99 \times 10^{-9} \), far below the standard alpha threshold of 0.05. This statistically significant result indicates a meaningful difference in user satisfaction ratings before and after repairs.
As seen in \cref{fig:u2}, the mean satisfaction score rose from 3.04 to 3.92 when applying repair using our work, with a notable shift in score distribution towards higher ratings. This upward trend, coupled with a reduction in standard deviation (from 1.19 to 0.99), suggests a more uniformly positive user experience following the repairs.

\vspace{-4mm}

\revision{Knowing the order of the videos, we calculated how often the before UI page scored lower, equal to, or higher than the after-repair UI page.}
As shown in \cref{fig:u3}, a notable 58.7\% of users regarded the pages as visually superior before-repair. \revision{In the interview, participants provided feedback indicating an enhanced user experience,} evidenced by greater ease of use, more efficient information retrieval, and a noticeable increase in design attractiveness.

Interestingly, 20.4\% of participants rated the before and after repair versions equally. Feedback from these participants\revision{'s interview} indicated that in cases where the frontend code was complex and the errors minor (involving aspects like content and accessibility features), the distinctions in repair quality were not immediately apparent.
Conversely, 20.1\% of participants expressed a preference for the original design.
\revision{After being informed about the guideline checks and repairs for readability and interactivity, some participants still preferred the original versions. They felt that certain soft constraint repairs were subjective. Moreover, some design issues, like colorful themes, did not affect usability and were more engaging.}
Overall, our approach demonstrates its effectiveness through both statistical and descriptive analyses. 

\subsection{\revision{Threats to Validity}} 
\revision{Due to the limitations of manually recognizing design issues in the code, the distribution of identified design issues may differ from the broader context. For example, annotators found it more challenging to identify soft constraints. Therefore, our statistics may not fully reflect the proportions in AI-generated and GitHub project issues. There may be some bias. Testing on a larger dataset in the future could give additional insights.}

\section{\revision{Discussion and Future Work}}

\revision{Our work presents an extensible, knowledge-based code repair method that effectively utilizes design guidelines without requiring manual rule building and development.} This approach's strength lies in its flexibility to support and integrate diverse guidelines.
By leveraging LLMs and an extendable knowledge-driven framework, our method can adapt seamlessly to evolving guidelines. 
\revision{Furthermore, future improvements could focus on mechanisms to effortlessly incorporate guideline updates and further narrow the scope of LLM inspection. We have identified a promising direction by combining related specifications with relevant code, properties, and values to achieve more effective performance. We found that shortening the reasoning process can enhance code repair capabilities. Combining LLMs with traditional page analysis tools can create a more efficient workflow by leveraging the unique strengths of each tools. 
These advancements ensure alignment with industry standards, enabling organizations to tailor the approach to their specific needs.}

\revision{Additionally, our work provides a first inspection of LLM-generated frontend code issues and methods to improve them, which is valuable for enhancing generated code, especially frontend code. In the future, we can incorporate LLM issue detection and repair during the code generation process, enabling early problem identification and more reliable code creation.}

\section{Related work}

\subsection{Accessibility Testing Tools}
Widely-used accessibility testing tools like Playwright~\cite{microsoft_playwright_2024}, Google Lighthouse~\cite{googlechrome_lighthouse_2024}, and axe-core~\cite{dequelabs_axe_core_2024} are primarily designed for addressing accessibility issues like text contrast, icon tapability, image accessibility, display issues, and intention-practice mismatches. 
These tools lack various aspects of repair capabilities and often miss many issues. A user study~\cite{ismailova2022comparison} found they detected only about 50\% of the issues they claimed, highlighting their limitations.

\subsection{Research for UI Issues}

To address the limitations of accessibility testing tools, researchers have developed tools targeting specific UI issues. OwlEye~\cite{liu2020owl}, IFIX~\cite{mahajan2018automated}, CBRepair~\cite{alameer2019efficiently}, and MFIX~\cite{mahajan2018automated} focus on Android GUI layout issues, web application internationalization presentation problems, and mobile web presentation issues, respectively. Iris~\cite{zhang2023automated} and Seenomaly~\cite{zhao2020seenomaly} are specifically designed for color-related accessibility issues and detecting violations of design guidelines in animations. While these tools have contributed to advancing QA in their respective domains, 
they lack general, adaptive, and effective methods to address diverse frontend design issues.

UIS-Hunter~\cite{yang2021don} represents a vital effort in design smell detection by encompassing 23 types of UI components against 126 visual guidelines. This extensive scope marks a notable advancement in identifying a wide array of design issues. Despite its broad coverage, UIS-Hunter struggles with complex design principles, focusing on component-level guidelines while missing higher-level aspects like layout issues.
Furthermore, it can only detect issues and lacks repair ability.

\subsection{LLM-based Code Generation and Repair}

The emergence of automated program repair research, coupled with the advancements in LLMs, opens up new possibilities for \revision{an open approach for} program repair. Pre-trained LLMs like CodenBERT~\cite{feng2020codebert}, CodeX~\cite{chen2021evaluating}, AlphaCode~\cite{li2022competition}, and Code Llama~\cite{r2024codellama} have demonstrated significant improvements in code generation and repair. Research on performance enhancement in program repair has also progressed, with models like CodeT5~\cite{wang2021codet5}, CodeGen~\cite{nijkamp2022conversational}, and InCoder~\cite{fried2022incoder} showing competitive or superior performance compared to deep learning-based techniques. These models leverage knowledge-driven methods to enhance repair capabilities without additional model training.

However, while those LLMs have shown remarkable progress in code generation and repair, their focus has primarily been on broader problems and languages, such as algorithms, logic problems, Python, and C. Advanced LLM methods like MetaGPT~\cite{hong2023metagpt} and KPC~\cite{ren2023misuse} have improved Python code project, and Java exception handling issue quality within the LLM QA framework, but frontend code repair using LLMs remains an underexplored area.

In summary, while accessibility testing tools, research tools for specific UI issues, they fail to provide a \revision{a broader spectrum and adaptable} solution for frontend code repair. The remarkable progress in LLM-based code generation and repair presents an opportunity to explore potential ways to address the extensive design issues of frontend for QA. Our work focuses on this critical unexplored domain, building upon the advancements in LLMs and program repair techniques.

\section{Conclusion} 

\revision{In this paper, we present \tool{}, a dual-stream, design guideline-aware system developed to provide a guidelines-adaptive method for detecting and repairing frontend code. To evaluate its effectiveness, we analyzed 20 projects created using Vercel's V0 and 66 files from 6 GitHub projects. Our manual examination revealed insights into the tool's adherence to design guidelines. We conducted evaluations and a user study to demonstrate its effectiveness.}



\bibliographystyle{IEEEtran}
\bibliography{refer}

\end{document}